\documentclass[showpacs,amsmath,amssymb,floatfix,prl,twocolumn]{revtex4}
\usepackage{graphicx,amsmath}

\begin{document}

\title { Magnetic field anti-symmetry of photovoltaic voltage in evanescent microwave fields}

\author{ A. Chepelianskii$^{(a)}$, S. Gu\'eron$^{(a)}$, F. Pierre$^{(b)}$, A. Cavanna$^{(b)}$,  B. Etienne$^{(b)}$ and H. Bouchiat$^{(a)}$ }
\affiliation{$(a)$ Univ. Paris-Sud, CNRS, UMR 8502, F-91405, Orsay, France }
\affiliation{$(b)$ Laboratoire de Photonique et de Nanostructures (LPN)-CNRS, route de Nozay, 91460 Marcoussis, France }

\begin{abstract}
A two dimensional electron system without spatial inversion symmetry develops
a sample specific dc voltage when exposed to a microwave radiation at low temperature. 
We investigate this photovoltaic (PV) effect, in the case where 
spatial symmetry is broken by an evanescent high-frequency potential. 
We measure the induced PV voltage in a ${\rm GaAs/Ga_{1-x}Al_{x}As}$ Hall bar 
at magnetic fields in the Tesla range. 
We find that in this regime the induced PV voltage is anti-symmetric with magnetic field,
and exhibits regular Shubnikov-de Haas like oscillations. Our experimental results 
can be understood from a simple model, which describes the effect of stationary 
orbital currents caused by microwave driving. 
\end{abstract}

\maketitle

\begin{figure}
\begin{center}
\vglue -0.5cm
\includegraphics[clip=true,width=8cm]{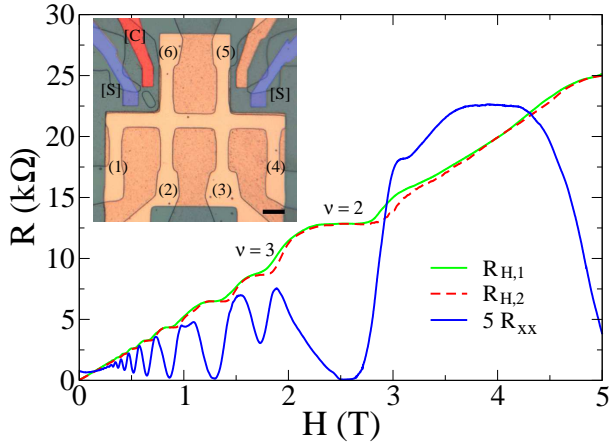}
\caption{ (Color online) Magneto-transport properties of the Hall bar. 
The dashed and neighbor curves represent respectively the Hall resistances $R_{H,2} = R_{14,35}$ and $R_{H,1} = R_{14,26}$ (see text). 
The oscillating curve represents the longitudinal resistance $R_{xx} = R_{14,23}$. 
The data was acquired with a $100$ nA excitation and lock-in detection at frequency $67$ Hz at a temperature of $300$ mK. 
The inset is an optical image of the sample, the leads $(1)-(6)$ form the contacts of the Hall bar, 
while the electrodes $[S]$ and $[C]$ are local gates connected to a high frequency $50 \;{\rm \Omega}$ transmission line. 
The (orange) copper shield on top of the Hall bar can be used as a top gate to change the carrier density in the 2DEG. 
The black scale bar corresponds to $10\;{\rm \mu m}$. 
\label{samplepanddIdVhighV}}
\end{center}
\end{figure}

Coherent mesoscopic samples present remarkable rectification properties,
related to the absence of spatial inversion symmetry of the disorder potential. 
In particular when submitted to a radio-frequency radiation they develop 
a dc voltage. The dependence on magnetic field of this photovoltaic (PV) effect, 
gives rise to random but reproducible fluctuations that were predicted theoretically \cite{altkm,falko}
and observed experimentally in references \cite{webb88,bykov}. 
In contrast with universal conductance fluctuations that obey Onsager 
symmetry rules \cite{webb,Buttiker}, the fluctuations of the PV-voltage do not have 
a well defined magnetic field symmetry \cite{sanchez}. At high frequency this can be 
understood from the violation of time inversion symmetry by the microwave radiation \cite{giordano,marcus,angers08}.
At low frequencies, this  behavior was explained through a mechanism involving 
electron-electron interactions \cite{wei05,leturq06,marcus06,angers06,monod,moskalets}.
Recently the PV effect was studied in asymmetric antidot super-lattices where 
magnetic field asymmetry was also present \cite{sassine,toulouse}.
In all these cases however, the anti-symmetric component of the PV voltage was never larger than the symmetric one
(see e.g. \cite{vitkalov}). 
In this Letter we investigate the regime where the spatial symmetry of the system is broken by 
a non homogeneous high frequency potential.
This potential is screened in the region of the sample where the photovoltaic voltage is measured
by a copper shield evaporated on the surface of the Hall bar.
This reduces the mesoscopic fluctuations of the PV-voltage, revealing  
a PV-voltage with a dominant anti-symmetric contribution. 

The system consists of two connected Hall junctions in a ${\rm GaAs/Ga_{1-x}Al_{x}As}$ two dimensional 
electron gas (2DEG) with density  $n_e \simeq 1.2 \times 10^{11}\;{\rm cm^{-2}}$ and mobility $\mu \simeq 1.1 \times 10^2\;{\rm m^2/Vs}$.
The sample was fabricated using wet etching and an aluminum mask. The six contacts of the Hall bar are numbered $(1)-(6)$ 
(see photograph inset in Fig.~1).
The four terminal resistances with source $(i)$ and drain $(j)$ and voltage probes $k$ and $l$ are defined by 
the usual relation $R_{ij,kl} = (V_k - V_l) / I_i$ where $V_k$ and $V_l$ are the voltages on the leads $(k)$ and $(l)$ 
and $I_i = -I_j$ is the current injected in the source lead. We measure simultaneously the Hall resistances 
$R_{H,1} = R_{14,26}$, $R_{H,2} = R_{14,35}$, and the longitudinal resistance $R_{xx} = R_{14,23}$ as a function 
of magnetic field $H$. 
As shown on Fig.~1 our samples exhibit quantum Hall effect plateaux \cite{Klitzing} for the Hall resistances $R_{H,1}, R_{H,2}$ 
and Shubnikov-de Haas (SdH) oscillations \cite{Shoenberg} in the transverse resistance $R_{xx}$ . We notice that the carrier density in our system is homogeneous 
since $R_{H,1} \simeq R_{H,2}$.

\begin{figure}
\begin{center}
\includegraphics[clip=true,width=8.5cm]{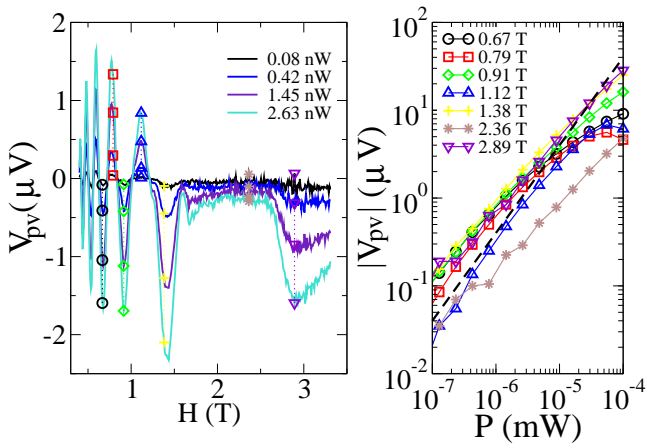}
\caption{(Color online) Left panel: Magnetic field dependence of the photovoltaic voltage $V_{pv}$ for different microwave powers, 
namely $0.08, 0.42, 1.45$ and $2.63$ nW in the direction of increasing oscillation amplitude. Right panel: 
Dependence of the photovoltage amplitude $V_{pv}$ as a function of microwave power for different values 
of the magnetic field (the symbols correspond to different magnetic fields and are indicated on both 
the right and left panel). The dashed line represents the dependence $V_{pv} \propto P$. 
Temperature is $300\;{\rm mK}$, and frequency $f = 2.5\;$GHz.
\label{scalingdynamicCB}}
\end{center}
\end{figure}   

In the following, we excite the system with a high frequency potential $V_{ac} \cos(2 \pi f t)$ 
applied on a symmetrical split gate $[S]$ shown in the inset of Fig.~1.  This potential 
is screened by the copper top-gate deposited over the Hall bar device and therefore the induced potential 
is spatially non homogeneous and vanishes exponentially inside the Hall bar \cite{stadium}. 
We measure the PV voltage drop $V_{pv} = V_{3} - V_{2}$ induced by the irradiation. 
In order to determine this voltage with a high precision we modulate the high 
frequency signal at a low frequency below $1.5$ kHz. The voltage $V_{pv}$ is then 
amplified by a low-noise amplifier and measured by a lock-in detector working at 
the frequency of the amplitude modulation. 
In Fig.~2, we have studied the dependence of $V_{pv}$ on magnetic field and microwave power at fixed 
frequency $f = 2.5$ GHz. The photovoltaic voltage displays oscillations as a function of 
magnetic field reminiscent of SdH oscillations in the longitudinal 
resistance $R_{xx}$ (Fig.~2 left panel), except that the PV voltage oscillates around a zero mean value which is not 
the case for the longitudinal resistance.
We also notice that the oscillations of $V_{pv}$ are quenched in the quantum Hall plateau region around $H = 2\;$T
(filling factor $\nu = 2$), in contrast to oscillations in longitudinal resistance (see Fig.~1).
The amplitude of these oscillations increases with the injected microwave power $P$.
Since the microwaves are transmitted through a $Z_0 = 50\; {\rm \Omega}$ adapted line, 
$P$ is related to the high frequency potential amplitude through $V_{ac}^2 = \alpha Z_0 P$. 
Here $\alpha$ is a frequency dependent coefficient taking into account the attenuation 
and reflection in the transmission line. We estimate that $\alpha \simeq 0.1$ for frequencies in the GHz range. 
On the right panel of Fig.~2 we show the dependence 
of the rectified voltage on power more quantitatively at selected values 
of magnetic field. We find that the photovoltaic voltage is well described by $V_{pv} \propto P$, 
the deviations at higher power are attributed to heating effects.  
We have also performed experiments with asymmetric irradiation on the electrode $[C]$, with the electrode opposite to $[C]$ floating. 
These experiments lead to a similar behavior  with the difference that the oscillations 
of the PV voltage are no longer centered around zero. For this reason we focus on the case of symmetric irradiation on 
the local gate $[S]$ in the rest of this article.

\begin{figure}
\begin{center} 
\includegraphics[trim=0mm 10mm 45mm 23mm,clip=true,width=9cm]{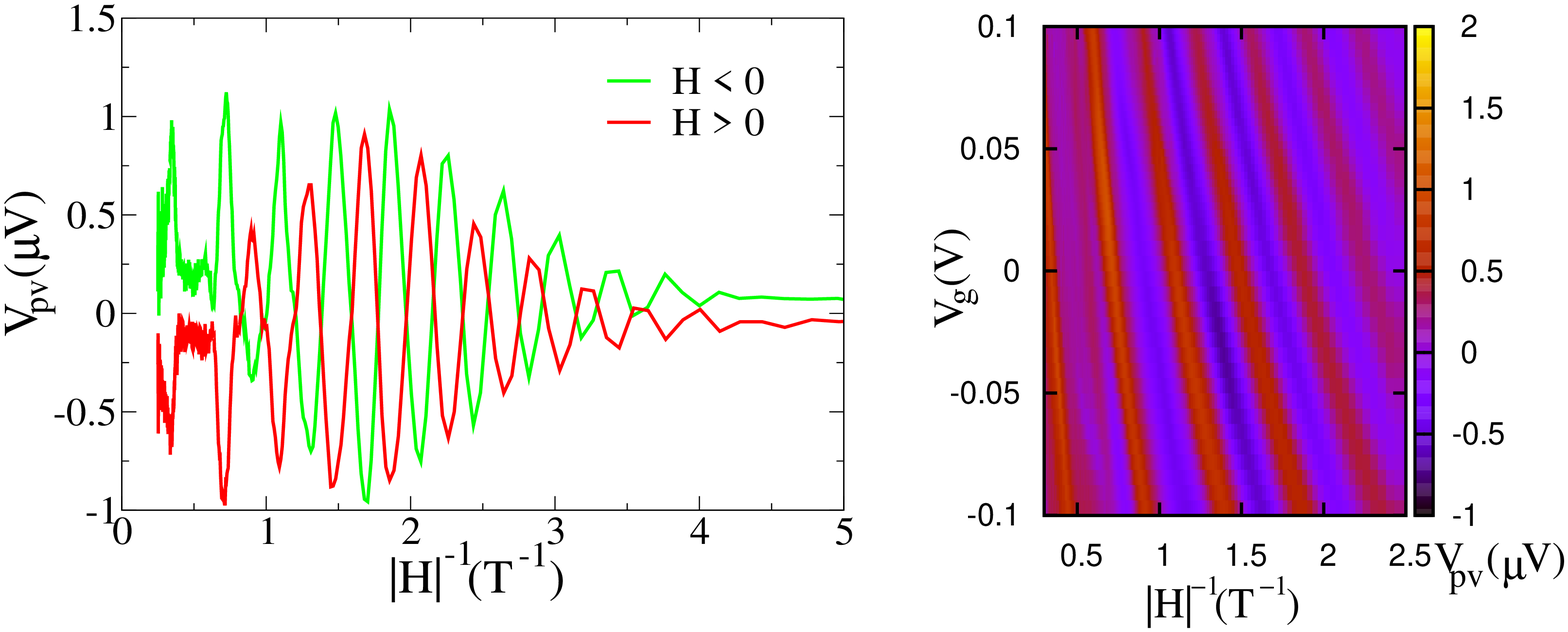}
\leavevmode
\caption{ (Color online) Left panel: Comparison of the PV voltage $V_{pv}$ as a function of inverse magnetic field for positive 
and negative magnetic fields. Frequency was $f = 2.5\;$GHz and injected microwave power of $1\;{\rm nW}$. 
On the right panel, a color diagram showing the PV voltage as a function of both top-gate voltage and applied magnetic field. 
\label{peaksg}}
\end{center}
\end{figure}

The Shubnikov-de Haas (SdH) oscillations of resistivity in metals and in 2DEG are symmetric with magnetic field.
In contrast as illustrated on the left panel of Fig.~3, 
we find that the oscillations of the photovoltage are mostly anti-symmetric with magnetic field.
On this figure, in order to make the connection with SdH oscillations more obvious 
we have shown the photovoltage $V_{pv}$ as a function of the inverse absolute value of the magnetic field. 
The photovoltage displays periodic oscillations with inverse magnetic field with the exception of a
missing half-period at $H^{-1} \simeq 0.5\;{\rm T^{-1}}$ in the quantum-Hall plateau regime (filling factor $\nu = 2$). 
The right panel shows that the period of the oscillations of the photovoltage can be modified 
by changing the density $n_e$ of the Hall probe with a top-gate voltage $V_g$. This further supports 
the connection with SdH oscillation of resistivity whose period $\tau_H$ is related to electron density 
through $\tau_H = e / (\pi \hbar n_e)$. We have also checked the anti-symmetry of the photovoltage at 
different gate voltages. 
The data shown on Fig.~3 was obtained for $f = 2.5$GHz, at lower microwave frequencies the 
photovoltage decreases and vanishes 
around $f \simeq 10$MHz while remaining mostly anti-symmetric. We attribute this decrease to the inefficiency 
of our capacitive coupling at so low frequencies. We note that we are always in the regime $2 \pi f < \omega_c$ where $\omega_c$ is
the cyclotron frequency, therefore we do not expect the frequency to play an important role in contrast 
with recent experiments where a sharp frequency dependent PV effect was investigated around $\omega = \omega_c$ \cite{bykov2}
in relation with microwave induced resistance oscillations and zero resistance states \cite{zudov1,mani,zudov2}. 

\begin{figure}
\begin{center}
\includegraphics[clip=true,width=8cm]{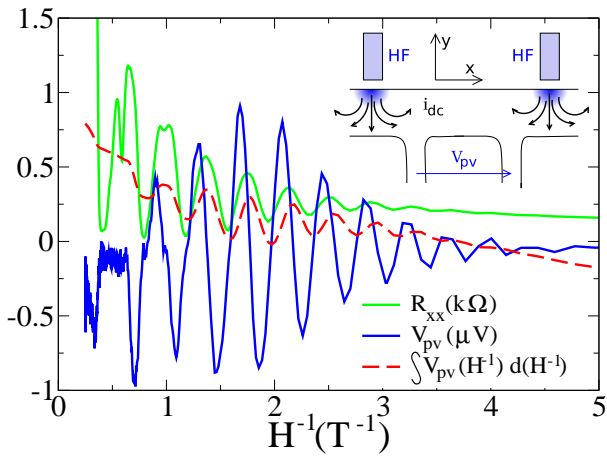}
\caption{(Color online) Comparison between the behavior of $R_{xx} \; ({\rm k \Omega})$ (green/gray curve), 
 $V_{pv} \; ({\rm \mu V})$ (blue/black curve) and $\int V_{pv}(H^{-1}) d H^{-1}$ (red dashed curve). 
The inset shows a simplified sample geometry with schematic representation of induced stationary currents $i_{dc}$, 
and density modulation (colored semi-disks). Frequency is $f = 2.5\;{\rm GHz}$ and $P = 1\;{\rm nW}$.
\label{magneticfield}}
\end{center}
\end{figure}

The SdH oscillations of longitudinal resistance $R_{xx}$ as a function of 
inverse magnetic field have a well defined phase, which is given by the expression
$R_{xx}(H^{-1}) = R_0(H^{-1}) + R_1(H^{-1}) \cos( \frac{2 \pi^2 n_e \hbar}{e H} )$, 
where $R_0(H^{-1})$ and $R_1(H^{-1})$ are envelope functions weakly depending on the magnetic field on the scale 
of one SdH oscillation period $\tau_H$ \cite{Shoenberg}. On Fig.~4 (main figure) we compare the phase of the oscillations of $V_{pv}$ 
and the phase of the oscillations in $R_{xx}$. Our results demonstrate that the oscillation of the photovoltage are dephased by $\pi/2$ 
compared to the oscillations of $R_{xx}$. We check this by calculating numerically the integral $\int V_{pv}(H^{-1}) d H^{-1}$ 
from the experimental data. We find that oscillations of this quantity are in phase with the oscillations of $R_{xx}(H^{-1})$ 
(a theoretical argument justifying this comparison is given below).

In summary we find that the SdH oscillations of the PV voltage are anti-symmetric with magnetic field 
and are out of phase with the usual SdH oscillation of resistivity as a function of $H^{-1}$. 
These oscillations are quenched in the plateau regions of the quantum Hall effect.
The main lines of our explanation are the following. The high frequency potential applied on the local gate $[S]$   
creates a stationary current distribution $i_{dc}$ inside the sample mostly along the y axis. 
A possible current distribution is shown in the inset 
of Fig.~4 with a simplified sample geometry. This current leads to the appearance of a static Hall voltage drop 
$V_{pv} \simeq R_{H} i_{dc}$ along the x axis perpendicular to $i_{dc}$. 
For this expression to be valid, the applied potential 
$V_{ac}$ must vanish in the region where $V_{pv}$ is measured. Otherwise an additional contribution 
appears from mesoscopic fluctuations induced by the alternating potential. This contribution has a large 
symmetric component. In this respect, the evanescent potential geometry used in our experiment is crucial, and 
we have checked that under an homogeneous irradiation we recover an essentially symmetric photovoltage. 

It is possible to derive more quantitative estimates from this 
heuristic scenario. The potential $V_{ac}$ on the local gate creates a modulation of the electronic density $\delta n$ in the Hall bar.
We assume that for high magnetic fields this density modulation occurs in a region of typical Larmor radius $r_l = v_F m^* / e H$ 
inside the sample, where $v_F = \hbar \sqrt{2 \pi n_e} / m^{*}$  is the Fermi velocity and $m^*$ the electron mass in 2DEG
(colored regions in the sketch of Fig.~4). 
Indeed in this regime most electronic trajectories are localized on cyclotron orbits of radius $r_l$,
which is therefore the natural length scale. 
The total charge induced on the Hall bar is given by $C V_{ac}$ where $C$ is the capacitance between the local gate  
and the Hall bar, hence the amplitude of $\delta n$ can be estimated as : $\delta n \simeq \frac{C}{r_l^2} V_{ac} \cos(2 \pi f t)$. 
An approximate value of the capacitance is $\epsilon D$, 
where $\epsilon$ is the permittivity  and $D$ is the typical distance between the gate and the Hall bar.
In the SdH regime, the amplitude of the rectified current $i_{dc}$ is given by: 
\begin{align}
i_{dc} \simeq <\frac{\partial G_{yy}}{\partial n_e} \delta n \; V_{ac} \cos(2 \pi f t)>
\end{align}
where $<.>$ denotes time averaging, and $G_{yy}$ is the conductance in $y$ direction of the sample. 
Up to a geometrical factor we have $G_{yy} \simeq G_{xx}$ 
and in the regime where $R_{xx} \ll R_{H}$, $G_{xx}$ is given by $G_{xx} \simeq R_{xx} / R_H^2$. 
As noted above, $R_{xx} = R_0 + R_1 \cos( \frac{2 \pi^2 n_e \hbar}{e H} )$ 
where $R_0$ and $R_1$ are slow functions of density and inverse magnetic field. This allows us 
to take into account only the oscillating term in the density derivative, which yields : 
$i_{dc} \simeq \frac{\hbar}{e H} \frac{ R_1 \sin( \frac{2 \pi^2 n_e \hbar}{e H} ) }{ R_H^2} <\delta n V_{ac} \cos(2 \pi f t)>$. 
By injecting in this formula the expression of $\delta n$ as a function of $V_{ac}$, and using the relation $V_{pv} = i_{dc} R_H$ 
with the approximation $R_H = H / (e n_e)$ that is accurate below $1\;{\rm T}$, we obtain 
\begin{align}
V_{pv} \simeq \frac{e C}{\hbar} R_1(H) \sin \left( \frac{2 \pi^2 n_e \hbar}{e H} \right) V_{ac}^2
\end{align}
In this expression $R_1(H)$ is the typical SdH oscillation amplitude that is symmetric 
with magnetic field. Consequently this expression reproduces the main features observed in our experiment: 
$V_{pv}$ is anti-symmetric with magnetic field, with a phase given by $\sin \left( \frac{2 \pi^2 n_e \hbar}{e H} \right)$,
and an amplitude that scales proportionally to microwave power $V_{pv} \propto V_{ac}^2 \propto P$.
This expression also leads to the right order of magnitude for the observed photovoltaic voltage, 
indeed for $C \simeq \epsilon_0 D$ with $D \simeq 1 \; {\rm \mu m}$, $V_{ac}^2 = 1\; {\rm nW} \times 50\;{\rm \Omega}$
 and $R_1(H) \simeq 1\; {\rm k \Omega}$ at $H \simeq 1\;{\rm T}$
we find that $V_{pv} \simeq 0.1\;{\rm \mu V}$ (experimental amplitude is shown on the right panel of Fig.~2).
Further comparison is possible by noting that the product $R_1(H) \sin \left( \frac{2 \pi^2 n_e \hbar}{e H} \right)$ 
is proportional to $d R_{xx}(H^{-1})/ d H^{-1}$. In our approximation, this leads to the simple prediction: 
\begin{align}
\int V_{pv}(H^{-1}) d H^{-1} \propto R_{xx}(H^{-1}) 
\end{align}
On figure Fig.~4, we show that this relation is well verified as long as $H^{-1} \ge 1.5\;{\rm T^{-1}}$. 
Deviations are observed for higher magnetic fields, specially at $H^{-1} \simeq 0.5\;{\rm T^{-1}}$ where a plateau appears in $V_{pv}$ that is 
not present in $R_{xx}$. We attribute this deviation to other quantum effects which are not taken into account in our 
simple model and become relevant at higher magnetic field. Specially the physics of the quantum Hall effect 
is important, while our model is based on a Shubnikov-de Haas approximation for conductivity. 
In order to confirm that the effect we have observed has a quantum origin, we have measured the temperature dependence 
of photovoltaic effect. Our results summarized on Fig.~5, show that the amplitude of the photovoltaic oscillations 
strongly decreases with temperature in the $100\;{\rm mK}$-$1\;{\rm K}$ range, in a similar way to
SdH oscillations on the linear resistance. 

\begin{figure}
\begin{center}
\includegraphics[clip=true,width=7cm]{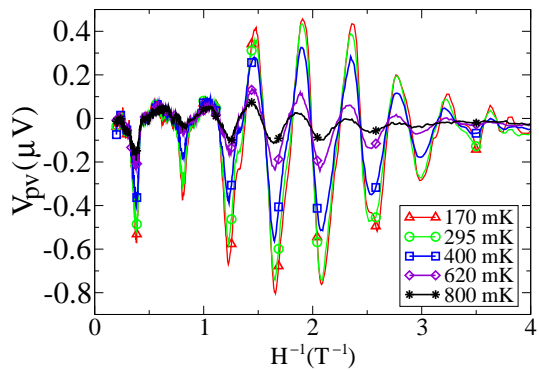}
\caption{(Color online) Photovoltage dependence on inverse magnetic fields for different temperatures for $f = 2.5\;{\rm GHz}$ and $P = 1\;{\rm nW}$.}
\end{center}
\end{figure}

In conclusion we have investigated the photovoltaic effect in high mobility two dimensional 
electron gas under irradiation by an evanescent microwave potential. 
We have found that the photovoltaic voltage exhibits zero centered oscillations as a function of inverse magnetic field. 
These oscillations are anti-symmetric with magnetic field and are out of phase with the well
known Shubnikov-de Haas oscillations in resistivity as a function of inverse magnetic field. 
The amplitude of these oscillations is proportional to microwave power. 
Our experimental findings can be understood from a simple model that predicts the creation of stationary 
orbital currents in the sample under microwave driving. 
In this model the stationary voltage across the sample appears as a Hall effect detection 
of the orbital currents. 

We are grateful to J. Gabelli, B. Reulet, M. Ferrier, R. Deblock and U. Gennser for stimulating discussions, 
and we acknowledge ANR NANOTERA and DGA for support.


\begin{thebibliography}{99}
\bibitem{stadium} We notice that the symmetry is broken by the presence of an oval dot between the left $[S]$ and $[C]$ gates, 
this weak asymmetry does not play a role for the effect discussed in this article but 
will be analyzed more carefully in a future publication.
\end{thebibliography}
\end{document}